\documentclass[a4paper]{amsart}

\usepackage[hypertexnames=false,linktocpage=true]{hyperref} 
\hypersetup{colorlinks=true,linkcolor=blue}

\usepackage{graphics,amssymb,enumerate}
\usepackage{hyperref}
\usepackage{graphicx}
\usepackage{float}
\floatstyle{boxed}
\restylefloat{figure}

\newcommand {\bd} {\begin{displaymath}}
\newcommand {\ed} {\end{displaymath}}
\newcommand {\be} {\begin{equation}}
\newcommand {\ee} {\end{equation}}
\newcommand {\bea} {\begin{eqnarray}}
\newcommand {\eea} {\end{eqnarray}}

\newcommand{\Ref}[1]{(\ref{#1})}

\newtheorem{example}{Example}

\DeclareMathOperator{\Rank}{Rank}

\newcommand{\MeR}{\\\noalign{\medskip}}
\newcommand{\LR}{\\\noalign{\bigskip}}
\newcommand{\NR}{\\\noalign{\medskip}}
\DeclareMathOperator{\tr}{tr}

\newcommand {\La} {\Lambda}
\newcommand{\Gp}{\ensuremath{\pi}}

\begin{document}
\title{On Generalized Volterra systems}

\author{  S. A. Charalambides, P. A. Damianou, C. A. Evripidou}
\address{Department of Mathematics and Statistics\\
University of Cyprus\\
P.O.~Box 20537, 1678 Nicosia\\Cyprus}
\email{charalambides.stelios@ucy.ac.cy, damianou@ucy.ac.cy, cevrip02@ucy.ac.cy}

\date{}

\begin{abstract}
We construct a large family of evidently integrable Hamiltonian systems which are generalizations of the KM system.   The Hamiltonian vector field is homogeneous cubic but in a number of cases a simple change of variables transforms such a system to a quadratic Lotka-Volterra system. We present in detail all such systems in dimensions 4 and 5 and we also give some examples from higher dimensions.  This construction generalizes easily to  each complex simple Lie algebra.
\end{abstract}

\maketitle

\section{Introduction} \label{intro}

The Volterra model, also known as the  KM system is a well-known integrable system defined by
\begin{equation} \label{a1}
\dot x_i = x_i(x_{i+1}-x_{i-1}) \qquad i=1,2, \dots,n,
\end{equation}
where $x_0 = x_{n+1}=0$. It was studied by
Lotka in \cite{lotka} to model oscillating chemical reactions and by
Volterra in \cite{volterra} to describe population evolution in a
hierarchical system of competing species. It was first solved by
Kac and van-Moerbeke in \cite{kac}, using a discrete version of
inverse scattering due to Flaschka \cite{flaschka}. In
\cite{moser} Moser gave a solution of the system using the method
of continued fractions and in the process he constructed
action-angle coordinates. Equations \Ref{a1} can be considered
as a finite-dimensional approximation of the Korteweg-de Vries
(KdV) equation.  The Poisson bracket  for this system can
be thought as a lattice generalization of the Virasoro algebra
\cite{fadeev2}.
 The Volterra system is associated with a simple Lie algebra of type $A_n$. Bogoyavlensky
generalized this system for each simple Lie algebra and showed
that the corresponding systems are also integrable. See
\cite{bog1,bog2} for more details. The generalization in this paper is different from the one of Bogoyavlensky.

The KM-system given by equation \Ref{a1} is Hamiltonian (see \cite{fadeev}, \cite{damianou91}) and can be written in Lax pair form in various ways.
The Lax pair in \cite{damianou91}  is given by
\begin{equation*}
\dot{L}=[B, L],
\end{equation*}
where
\begin{equation*}
  L= \begin{pmatrix} x_1 & 0 & \sqrt{x_1 x_2} & 0 & \cdots &   & 0 \cr
0 & x_1 +x_2 & 0& \sqrt{x_2 x_3}  & & & \vdots \cr
 \sqrt{x_1 x_2} & 0 & x_2 +x_3 & &  \ddots & &  \cr
 0 & \sqrt{x_2 x_3} & &  & & & \cr
  \vdots & & \ddots & & & & \sqrt{x_{n-1} x_n} \cr
 & & & & & x_{n-1}+x_n & 0 \cr
0 &  & \cdots & & \sqrt{x_{n-1} x_n} &0& x_n  \end{pmatrix}
 \end{equation*}
and
\begin{equation*}
 B=\begin{pmatrix}  0 & 0 & \frac{1}{2} \sqrt{x_1 x_2} & 0 & \dots &
\ & 0 \cr
 0 & 0 & 0&\frac{1}{2} \sqrt{x_2 x_3}  & & & \vdots \cr
 -\frac{1}{2} \sqrt{x_1 x_2} & 0 & 0 & &  \ddots & &  \cr
 0 & -\frac{1}{2} \sqrt{x_2 x_3} & &  & & & \cr
  \vdots & & \ddots & & & & \frac{1}{2} \sqrt{x_{n-1} x_n} \cr
 & & & & & 0 & 0 \cr
 0& & \cdots & &-\frac{1}{2}\sqrt{x_{n-1} x_n} &0& 0  \end{pmatrix} \ .
 \end{equation*}

Due to the Lax pair, it  follows that the functions	$H_i=\frac{1}{i}\, \tr \, L^i$ are constants of motion.
Following \cite{damianou91} we define the following quadratic
Poisson bracket,

 \begin{displaymath}  \{x_i, x_{i+1} \}=x_i x_{i+1}, \end{displaymath}
  and all
other brackets equal to zero.
This bracket has a single Casimir  det$L$,  and  the functions $H_i$ are all in involution.
 Taking the function $\sum_{i=1} ^{n} x_i $ as the
Hamiltonian we obtain equations (\ref{a1}). This bracket can be
realized from the second Poisson bracket of the Toda lattice by
setting the momentum variables equal to zero \cite{fadeev}.

There is another Lax pair  where $L$ is in the nilpotent subalgebra  corresponding to the negative
roots.   The Lax pair is of the form  $ \dot{L}=[L, B] $ where

\begin{equation} 
\label{lax-volterra}
L= \begin{pmatrix} 0 &  1 & 0 & \cdots & \cdots & 0 \cr
                   x_1 & 0 & 1 & \ddots &    & \vdots \cr
                   0 & x_2 & 0 & \ddots &  &  \vdots \cr
                   \vdots & \ddots & \ddots & \ddots &\ddots & 0 \cr
                   \vdots & & & \ddots & \ddots & 1 \cr
                   0 & \cdots &  \cdots & 0 & x_{n} & 0  \end{pmatrix},
\end{equation}
and
\begin{equation*}
 B=\begin{pmatrix} 0 &  1 & 0 & \cdots & \cdots & 0 \cr
                   0 & 0 & 1 & \ddots &    & \vdots \cr
                   x_1 x_2 & 0 & 0 & \ddots &  &  \vdots \cr
                   \vdots & x_2 x_3 & \ddots & \ddots & & 0 \cr
                   \vdots & & & \ddots & \ddots & 1 \cr
                   0 & \cdots &  \cdots & x_{n-1} x_n & 0 & 0  \end{pmatrix}.
 \end{equation*}

 Finally, there is a symmetric version due to Moser where
\begin{equation} \label{lax-symmetric}
L= \begin{pmatrix} 0 &  a_1 & 0 & \cdots & \cdots & 0 \cr
                   a_1 & 0 & a_2 & \ddots &    & \vdots \cr
                   0 & a_2 & 0 & \ddots &  &  \vdots \cr
                   \vdots & \ddots & \ddots & \ddots & & 0 \cr
                   \vdots & & & \ddots & \ddots & a_n \cr
                   0 & \cdots &  \cdots & 0 & a_{n} & 0  \end{pmatrix},
\end{equation}
and
\begin{equation*}
 B=\begin{pmatrix} 0 & 0  & a_1 a_2 & \cdots & \cdots & 0 \cr
                   0 & 0 &  0 & \ddots &    & \vdots \cr
                   -a_1 a_2 & 0 & 0 & \ddots & a_2 a_3 &  \vdots \cr
                   \vdots & -a_2 a_3 & \ddots & \ddots & & a_{n-1} a_n \cr
                   \vdots & & & \ddots & \ddots & 0 \cr
                   0 & \cdots &  \cdots & -a_{n-1} a_n & 0 & 0  \end{pmatrix}.
\end{equation*}
\noindent
The change of variables $x_i=2a_i^2$ gives equations \Ref{a1}.

It is evident from the form of $L$ in the various Lax pairs, that the position of the variables $a_i$ corresponds to the simple root vectors of a root system of type $A_n$.  On the other hand a non-zero entry of the matrix $B$  occurs at a position corresponding to the sum of two simple roots $\alpha_i$ and $\alpha_j$. In this paper we generalize the Lax pair of Moser \Ref{lax-symmetric} as follows.

Instead of considering the set of simple roots $\Pi$, we begin with a subset $\Phi$ of the positive roots $\Delta^{+}$ which contains $\Pi$, i.e. $\Pi \subseteq \Phi \subseteq \Delta^{+}$.  For each such choice of a set $\Phi$ we produce (almost always) a Lax pair and thus a new Hamiltonian system.  In this paper we consider some specific examples for the case of $A_n$. In dimension 3 this procedure produces only two systems, the KM system and the periodic KM system. In dimensions 4 and 5 we introduce and study some new systems. We show that all such systems are Liouville  integrable. To establish integrability we use standard techniques  of Lax pairs and Poisson geometry, the method of chopping and  also a  particular technique of Moser which uses the square of the Lax matrix.  A number of these systems also  exist  in the form (\ref{lax-volterra}). In that case one can use the method of chopping to show integrability.

\section{Lotka-Volterra Systems}\label{lotka}

 The KM-system belongs to a large class of the so called Lotka-Volterra systems. The most
general form of the Lotka-Volterra equations is
$$\dot x_i = \varepsilon_i x_i + \sum_{j=1}^n a_{ij} x_i x_j, \ \ i=1,2, \dots , n.$$
 We may assume that there are no  linear terms ($\varepsilon_i=0$). We  also assume
that the  matrix $A= (a_{ij})$ is  skew-symmetric. All these systems can be written in Hamiltonian form  using the  Hamiltonian function
\begin{displaymath} H=x_1+ x_2 +\cdots + x_n \, .\end{displaymath}
Hamilton's equations take the form $\dot x_i = \{x_i, H\}=\sum_{j=1}^n\Gp_{ij}$ with quadratic functions \begin{equation} \label{quad} \Gp_{i,j}=\{ x_i,x_j\} = a_{ij} x_i x_j, \ \ i,j = 1,2, \dots , n. \end{equation}
From the skew symmetry of the matrix $A=(a_{ij})$ it follows that the Schouten-Nijenhuis bracket $[\Gp,\Gp]$ vanishes:
\[
\begin{split}
[\Gp,\Gp]_{ij}&=
2\left(a_{ij}\{x_ix_j,x_k\}+a_{jk}\{x_jx_k,x_i\}+a_{ki}\{x_kx_i,x_j\}\right)\\
&=2\left(a_{ij}(a_{jk}+a_{ik})+a_{jk}(a_{ki}+a_{ji})+a_{ki}(a_{ij}+a_{kj})\right)x_ix_jx_k=0\,.\\
\end{split}
\]
The  bivector field $\Gp$ is an example of a  \textit{diagonal Poisson structure}.

The Poisson tensor \Ref{quad} is Poisson isomorphic to the constant Poisson structure defined by the constant matrix $A$, see \cite{fairen}.  If $\mathbf{k}=(k_1, k_2 \cdots, k_n)$ is a vector in the kernel of $A$ then the function
\begin{equation*}
f=x_1^{k_1} x_2^{k_2} \cdots x_n^{k_n}
\end{equation*}
is a Casimir. Indeed for an arbitrary function $g$ the Poisson bracket $\{f,g\}$ is
\[
\{f,g\}=\sum_{i,j=1}^n\{x_i,x_j\}\frac{\partial f}{\partial x_i}\frac{\partial g}{\partial x_j}=\sum_{j=1}^n\left(\sum_{i=1}^na_{ij}k_i\right)x_jf\frac{\partial g}{\partial x_j}=0\,.
\]
If the matrix $A$ has rank $r$ then there are $n-r$ functionally independent  Casimirs. This type of integral can be traced back to Volterra \cite{volterra};  see also \cite{plank},  \cite{fairen}, \cite{bogo3}.

\section{Simple Lie algebras} \label{liealgebras}
\label{Procedure}
We recall the following procedure  from \cite{damianou12}.
Let $\mathfrak{g}$ be any simple Lie algebra equipped with its Killing form $\langle\cdot\,\vert\,\cdot\rangle$. One chooses
a Cartan subalgebra $\mathfrak{h}$ of $\mathfrak{g}$, and a basis $\Pi$ of simple roots for the root system $\Delta$ of $\mathfrak{h}$ in
$\mathfrak{g}$. The corresponding set of positive roots is denoted by $\Delta^+$.  To each positive root $\alpha$ one can
associate a triple $(X_\alpha,X_{-\alpha},H_{\alpha})$ of vectors in $\mathfrak{g}$ which generate a Lie subalgebra
isomorphic to $sl_2(\mathbf{C})$. The set $(X_\alpha, X_{-\alpha})_{\alpha \in \Delta^+}\cup (H_\alpha)_{\alpha \in
\Pi}$ is a basis of $\mathfrak{g}$, called a root basis.
Let  $\Pi=\{ \alpha_1, \dots, \alpha_{\ell} \}$ and let    $X_{\alpha_1}, \ldots, X_{\alpha_\ell}$ be the corresponding root vectors in  $\mathfrak{g}$.  Define
\begin{displaymath} L=\sum_{\alpha_i \in \Pi} a_i (X_{\alpha_i}+X_{-\alpha_i})   \ . \end{displaymath}
To find the matrix $B$ we use the following procedure. For each $i,j$  form the vectors
$\left[X_{\alpha_i},X_{\alpha_j}\right]$.  If $\alpha_i+\alpha_j $ is a root then
 include a term of the form $a_i a_j \left[X_{\alpha_i},X_{\alpha_j}\right]$ in $B$.
We make $B$ skew-symmetric by including the corresponding negative  root vectors $a_i a_j [X_{-\alpha_i},X_{-\alpha_j}]$. Finally, we define the system using the Lax pair
\begin{displaymath} \dot{L}=[L, B]  \ . \end{displaymath}
For a root system of type $A_n$ we obtain the KM system.

In this paper we generalize this algorithm as follows.  Consider a subset $\Phi$ of
$\Delta^{+}$ such that
\bd 
\Pi \subset \Phi \subset \Delta^{+} \ .
\ed

The Lax matrix is easy to construct
\begin{displaymath} 
L=\sum_{\alpha_i \in \Phi} a_i (X_{\alpha_i}+X_{-\alpha_i})   \ . 
\end{displaymath}

Here we use the following enumeration of $\Phi$ which we assume to have $m$ elements.  The variables $a_j$ correspond to the simple roots $\alpha_j$ for $j=1,2, \dots, \ell$.  We assign the variables $a_j$  for $j=\ell+1, \ell+2, \dots, m $ to the remaining roots in $\Phi$.
To construct the matrix $B$ we use the following algorithm.  Consider the set $\Phi \cup \Phi^{-}$ which consists of all the roots in $\Phi$ together with their negatives.  Let
\bd \Psi =\left\{ \alpha+\beta \ | \ \alpha,  \beta \in \Phi \cup \Phi^{-},  \alpha+\beta \in \Delta^{+} \right\} \ . \ed
Define
\bd B=\sum c_{ij} a_i a_j (X_{\alpha_i+\alpha_j}+X_{-\alpha_i - \alpha_j} )   \ed
where $c_{ij}=\pm 1$ if  $\alpha_i+\alpha_j \in \Psi$ with $\alpha_i,\alpha_j\in\Phi\cup\Phi^-$ and $0$ otherwise.  In almost all cases  we are able to make the proper choices of the sign of the $c_{ij}$ so that we can produce a Lax pair. For example   we are able to do this in all eight cases in $A_3$  and in all but five of the sixty four cases in $A_4$. In this paper we restrict our attention to the $A_n$ case. Examples from other Lie algebras will be presented in a future publication.

 \section{Examples in $A_3$ and $A_4$} \label{examples34}

\begin{example}\label{example A_3 root system} ($A_3$  root system)\\*
Let $E$ be the hyperplane  of $\mathbb{R}^4$ for which the coordinates sum to $0$ (i.e. vectors orthogonal to $(1,1,1,1)$). Let $\Delta$ be the set of vectors in $E$ of length $\sqrt{2}$ with integer coordinates.  There are $12$ such vectors in all.  We use the standard inner product in $\mathbb{R}^4$ and the standard orthonormal basis $\{ \epsilon_1, \epsilon_2, \epsilon_3,  \epsilon_4 \}$. Then, it is easy to see that $\Delta = \{ \epsilon_i-\epsilon_j \ | \ i \not= j \}$.  The vectors
\begin{displaymath}
 \begin{array}{lcl}
\alpha_1 & =& \epsilon_1 -\epsilon_2 \\
\alpha_2 & =& \epsilon_2 -\epsilon_3 \\
\alpha_3 & =& \epsilon_3 -\epsilon_4 \
\end{array}
\end{displaymath}
form a basis of the root system in the sense that each vector in $\Delta$ is a linear combination of these three vectors with integer coefficients,  either all nonnegative or all nonpositive. For example,  $\epsilon_1 -\epsilon_3=\alpha_1+\alpha_2$, $ \epsilon_2 -\epsilon_4 =\alpha_2+\alpha_3$ and $\epsilon_1-\epsilon_4=\alpha_1+\alpha_2+\alpha_3$.  Therefore $\Pi=\{\alpha_1, \alpha_2, \alpha_3 \}$,  and the set of positive roots $\Delta^{+}$  is given by
\begin{displaymath}
\Delta^{+}= \{ \alpha_1, \alpha_2, \alpha_3, \alpha_1+\alpha_2, \alpha_2+\alpha_3, \alpha_1+\alpha_2+\alpha_3  \} \ .
\end{displaymath}
If we take $\Phi=\{ \alpha_1, \alpha_2, \alpha_3, \alpha_1 +\alpha_2 \} $
then
$$\Phi \cup \Phi^-=\{ \alpha_1, \alpha_2, \alpha_3, \alpha_1 +\alpha_2 , -\alpha_1, -\alpha_2, -\alpha_3, -\alpha_1 -\alpha_2 \}
$$
and
$\Psi =\{ \alpha_1, \alpha_2,  \alpha_1 +\alpha_2 , \alpha_2+\alpha_3, \alpha_1+\alpha_2+\alpha_3 \}$.  In this example the variables $a_i$ for $i=1,2,3$ correspond to the three simple roots $\alpha_1, \alpha_2, \alpha_3$.  We associate the variable $a_4$ to the root $\alpha_1 +\alpha_2$.
We obtain the following Lax pair:
\label{ex1}
\[
L=  \begin {pmatrix}
0      &a_{{1}}&a_{{4}}&0      \NR
a_1    &0      &a_{{2}}&0      \NR
a_{{4}}&a_{{2}}& 0     &a_{{3}}\NR
0      &0      &a_{{3}}&  0
\end {pmatrix}
\]
\[
B=\begin {pmatrix}
0      &-a_4a_2&a_1a_2 &-a_4a_3\NR
a_4a_2 &0      &-a_1a_4&a_2a_3 \NR
-a_1a_2&a_1a_4 &0      &0      \NR
a_4a_3 &-a_2a_3&0      &0
\end {pmatrix}.
\]
 The Lax pair is equivalent to the following equations of motion:
\[
\begin{split}
\dot{a_1} & = a_1a^2_2-a_1a^2_4,         \\
\dot{a_2} & =-a_2a^2_1+a_2a^2_3+a_2a^2_4,\\
\dot{a_3} & =-a_3a^2_2+a_3a^2_4,         \\
\dot{a_4} & = a_4a^2_1-a_4a^2_2-a_4a^2_3 \, .
 \end{split}
\]
With the substitution $x_i=a_i^2$ followed by scaling  we obtain the following Lotka-Volterra system.
\[
\begin{split}
\dot{x_1} &=x_1x_2-x_1x_4,\\
\dot{x_2}&=-x_2x_1+x_2x_3+x_2x_4,\\
\dot{x_3}&=-x_3x_2+x_3x_4,\\
\dot{x_4}&=x_4x_1-x_4x_2-x_4x_3 \,.
\end{split}
\]
The system is integrable. There exist two functionally independent Casimir functions
$F_1=x_1 x_3={\rm det} \, L $ and $F_2=x_1 x_2 x_4$.
The additional integral is the Hamiltonian $H=x_1+x_2+x_3+x_4=\tr L^2$.

The standard quadratic Poisson bracket is given by
\[\pi=
\begin {pmatrix}
   0    & x_1x_2  &    0    & -x_1x_4 \MeR
-x_2x_1 &   0     & x_2x_3  & x_2x_4  \MeR
   0    & -x_3x_2 &    0    & x_3x_4  \MeR
x_4x_1  & -x_4x_2 & -x_4x_3 &   0
\end {pmatrix}.
\]
We can find the Casimirs by computing the kernel  of the matrix
\begin{displaymath}
A=\begin{pmatrix}
0 & 1& 0 & -1 \cr
-1& 0& 1& 1 \cr
0&-1&0&1 \cr
1&-1&-1&0
\end{pmatrix}.
\end{displaymath}
The two eigenvectors with eigenvalue $0$ are $(1,0,1,0)$ and $(1,1, 0, 1)$.  We obtain the two Casimirs $F_1=x_1^1 x_2^0 x_3^1 x_4^0=x_1 x_3$ and $F_2=x_1^1 x_2^1 x_3^0 x_4^1=x_1 x_2 x_4$.
\end{example}

There is a similar Lax pair defined  by the matrix
\[
L= \left( \begin {array}{cccc}
0&a_{{1}}&0&0\NR
a_{{1}}&0&a_{{2}}&a_{{4}}\NR
0&a_{{2}}&0&a_{{3}}\NR
0&a_{{4}}&a_{{3}}&0
\end {array} \right)
\]
but the resulting system  is isomorphic to the previous example.

The Lax pair $L,B$ corresponding to $\Phi=\{\alpha_1,\alpha_2,\alpha_3,\alpha_1+\alpha_2+\alpha_3\}$ is
\[ L=
 \left( \begin {array}{cccc}
0&a_{{1}}&0&a_{{4}}\NR
a_{{1}}&0&a_{{2}}&0\NR
0&a_{{2}}&0&a_{{3}}\NR
a_{{4}}&0&a_{{3}}&0
\end {array} \right),
\]
\[B=\left( \begin {array}{cccc}
0&0&a_{{1}}a_{{2}}-a_{{4}}a_{{3}}&0\NR
0&0&0&-a_{{1}}a_{{4}}+a_{{2}}a_{{3}}\NR
-a_{{1}}a_{{2}}+a_{{4}}a_{{3}}&0&0&0\NR
0&a_{{1}}a_{{4}}-a_{{2}}a_{{3}}&0&0
\end {array} \right).
\]
Using  the substitution $x_i=2a_i^2$  we obtain the periodic KM-system
\[
\begin{split}
\dot{x_1} &=x_1x_2-x_1x_4,\\
\dot{x_2}&=-x_2x_1+x_2x_3,\\
\dot{x_3}&=x_3x_4-x_3x_2,\\
\dot{x_4}&=x_4x_1-x_4x_3\,.
\end{split}
\]
The Poisson matrix is
\[\pi= \left( \begin {array}{cccc}
0&x_{{1}}x_{{2}}&0&-x_{{1}}x_{{4}}\NR
-x_{{1}}x_{{2}}&0&x_{{2}}x_{{3}}&0\NR
0&-x_{{2}}x_{{3}}&0&x_{{3}}x_{{4}}\NR
x_{{1}}x_{{4}}&0&-x_{{3}}x_{{4}}&0
\end {array} \right)
\]
with
$\Rank(\pi)=2$.

\noindent
In addition to the Hamiltonian
\bd
H=x_1+x_2+x_3+x_4
\ed
it possesses  two Casimirs $C_1=x_1 x_3$ and $C_2=x_2 x_4$.

\begin{example}\mbox{}\\*
The Lax equation $\dot{L}=[B,L]$, corresponding to
$\Phi=\{\alpha_1,\alpha_2,\alpha_3,\alpha_1+\alpha_2,\alpha_2+\alpha_3\}$
with

\[ L=
\left( \begin {array}{cccc}
0&a_{{1}}&a_{{4}}&0\NR
a_{{1}}&0&a_{{2}}&a_{{5}}\NR
a_{{4}}&a_{{2}}&0&a_{{3}}\NR
0&a_{{5}}&a_{{3}}&0\end {array} \right)
\]
and
\[B=\left( \begin {array}{cccc}
0&-a_{{4}}a_{{2}}&a_{{1}}a_{{2}}&-a_{{1}}a_{{5}}-a_{{4}}a_{{3}}\NR
a_{{4}}a_{{2}}&0&-a_{{1}}a_{{4}}-a_{{5}}a_{{3}}&a_{{2}}a_{{3}}\NR
-a_{{1}}a_{{2}}&a_{{1}}a_{{4}}+a_{{5}}a_{{3}}&0&-a_{{2}}a_{{5}}\NR
a_{{1}}a_{{5}}+a_{{4}}a_{{3}}&-a_{{2}}a_{{3}}&a_{{2}}a_{{5}}&0
\end {array} \right)
\]
 is equivalent to the following equations of motion
\[
\begin{split}
\dot{a_1} &=a_1a^2_2-a_1a^2_5-a_1a^2_4-2a_3a_4a_5,\\
\dot{a_2}&=a_2a^2_4+a_2a^2_3-a_2a^2_1-a_2a^2_5,\\
\dot{a_3}&=a_3a^2_5+a_3a^2_4-a_3a^2_2+2a_1a_4a_5,\\
\dot{a_4}&=a_4a^2_1-a_4a^2_2-a_4a^2_3,\\
\dot{a_5}&=a_5a^2_1-a_5a^2_3+a_5a^2_2.
\end{split}
\]
Note that the system is not Lotka-Volterra.
It is Hamiltonian with Hamiltonian function
$H=\frac{1}{2}\left(a_1^2+a_2^2+a_3^2+a_4^2+a_5^2\right)$.
The system has Poisson matrix
\[
\Gp=\left( \begin {array}{ccccc}
0&a_{{1}}a_{{2}}&-2\,a_{{4}}a_{{5}}&-a_{{1}}a_{{4}}&-a_{{1}}a_{{5}}\NR
-a_{{1}}a_{{2}}&0&a_{{2}}a_{{3}}&a_{{2}}a_{{4}}&-a_{{2}}a_{{5}}\NR
2\,a_{{4}}a_{{5}}&-a_{{2}}a_{{3}}&0&a_{{3}}a_{{4}}&a_{{3}}a_{{5}}\NR
a_{{1}}a_{{4}}&-a_{{2}}a_{{4}}&-a_{{3}}a_{{4}}&0&0\NR
a_{{1}}a_{{5}}&a_{{2}}a_{{5}}&-a_{{3}}a_{{5}}&0&0
\end {array} \right)
\]
of rank 4 . The determinant $C=(a_1a_3-a_4a_5)^2$ of $L$ is the Casimir of the system.
The trace of $L^3$ gives the additional constant of motion
$$F=\frac{1}{6}\tr\left(L^3\right)=a_1 a_2 a_4+ a_2 a_3 a_5 $$
and therefore the system is Liouville integrable.
\end{example}

\begin{example}\mbox{}\\*
Let
\[ L=   \left( \begin {array}{cccc}
0&a_{{1}}&0&a_{{5}}\NR
a_{{1}}&0&a_{{2}}&a_{{4}}\NR
0&a_{{2}}&0&a_{{3}}\NR
a_{{5}}&a_{{4}}&a_{{3}}&0
\end {array} \right)
\]
and
\[B=\left( \begin {array}{cccc}
0&a_{{4}}a_{{5}}&a_{{1}}a_{{2}}+a_{{5}}a_{{3}}&-a_{{1}}a_{{4}}\NR
-a_{{4}}a_{{5}}&0&-a_{{4}}a_{{3}}&a_{{1}}a_{{5}}+a_{{2}}a_{{3}}\NR
-a_{{1}}a_{{2}}-a_{{5}}a_{{3}}&a_{{4}}a_{{3}}&0&-a_{{4}}a_{{2}}\NR
a_{{1}}a_{{4}}&-a_{{1}}a_{{5}}-a_{{2}}a_{{3}}&a_{{4}}a_{{2}}&0
\end {array} \right)  \ .
\]
 The Lax pair is equivalent to the following equations of motion:
\[
\begin{split}
\dot{a_1} &=a_1a^2_5+a_1a^2_2-a_1a^2_4+2a_2a_3a_5,\\
\dot{a_2}&=a_2a^2_3-a_2a^2_1-a_2a^2_4,\\
\dot{a_3}&=-a_3a^2_5+a_3a^2_4-a_3a^2_2-2a_1a_2a_5,\\
\dot{a_4}&=-a_4a^2_5-a_4a^2_3+a_4a^2_1+a_4a^2_2,\\
\dot{a_5}&=-a_5a^2_1+a_5a^2_4+a_5a^2_3.
\end{split}
\]
The Poisson matrix is
\[
\Gp= \left( \begin {array}{ccccc}
0&a_{{1}}a_{{2}}&2\,a_{{2}}a_{{5}}&-a_{{1}}a_{{4}}&a_{{1}}a_{{5}}\NR
-a_{{1}}a_{{2}}&0&a_{{2}}a_{{3}}&-a_{{2}}a_{{4}}&0\NR
-2\,a_{{2}}a_{{5}}&-a_{{2}}a_{{3}}&0&a_{{3}}a_{{4}}&-a_{{3}}a_{{5}}\NR
a_{{1}}a_{{4}}&a_{{2}}a_{{4}}&-a_{{3}}a_{{4}}&0&-a_{{4}}a_{{5}}\NR
-a_{{1}}a_{{5}}&0&a_{{3}}a_{{5}}&a_{{4}}a_{{5}}&0
\end {array} \right)
\]
with
$ \Rank(\Gp)=4 $.

\noindent
The constants of motion are
\[
\begin{split}
H&=\frac{1}{2}\left(a_1^2+a_2^2+a_3^2+a_4^2+a_5^2\right)\;\text{(Hamiltonian)},\\
F&=\frac{1}{6}\tr\left(L^3\right)=a_{{1}}a_{{4}}a_{{5}}+a_{{2}}a_{{3}}a_{{4}},\\
C&=\det(L)=\left(a_1a_3-a_2a_5\right)^2\;\text{(Casimir)}.
\end{split}
\]
\end{example}

\begin{example}\mbox{}\\*
For the root system of type $A_4$ the Lax pair corresponding to
$$\Phi=\{\alpha_1,\alpha_2,\alpha_3,\alpha_4,\alpha_2+\alpha_3\}$$
is given by the matrices
\[ L=   \left( \begin {array}{ccccc} 0&a_{{1}}&0&0&0\NR
a_{{1}}&0&a_{{2}}&a_{{5}}&0\NR
0&a_{{2}}&0&a_{{3}}&0\NR
0&a_{{5}}&a_{{3}}&0&a_{{4}}\NR
0&0&0&a_{{4}}&0
\end {array} \right)
\]
and
\[B=\left( \begin {array}{ccccc}
0&0&a_{{1}}a_{{2}}&-a_{{1}}a_{{5}}&0\NR
0&0&-a_{{5}}a_{{3}}&a_{{2}}a_{{3}}&-a_{{5}}a_{{4}}\NR
-a_{{1}}a_{{2}}&a_{{5}}a_{{3}}&0&-a_{{2}}a_{{5}}&a_{{3}}a_{{4}}\NR
a_{{1}}a_{{5}}&-a_{{2}}a_{{3}}&a_{{2}}a_{{5}}&0&0\NR
0&a_{{5}}a_{{4}}&-a_{{3}}a_{{4}}&0&0
\end {array} \right).
\]
Using  the change of variables $x_i=2a_i^2$ the corresponding system becomes

\[
\begin{split}
\dot{x_1} &=x_1x_2-x_1x_5,\\
\dot{x_2}&=-x_2x_5+x_2x_3-x_2x_1,\\
\dot{x_3}&=x_3x_5+x_3x_4-x_3x_2,\\
\dot{x_4}&=x_4x_5-x_4x_3,\\
\dot{x_5}&=-x_5x_4-x_5x_3+x_5x_1+x_5x_2 \,.
\end{split}
\]
The Poisson matrix is
\[
\Gp= \left( \begin {array}{ccccc} 0&x_{{1}}x_{{2}}&0&0&-x_{{1}}x_{{5}}\NR
-x_{{1}}x_{{2}}&0&x_{{2}}x_{{3}}&0&-x_{{2}}x_{{5}}\NR
0&-x_{{2}}x_{{3}}&0&x_{{3}}x_{{4}}&x_{{3}}x_{{5}}\NR
0&0&-x_{{3}}x_{{4}}&0&x_{{4}}x_{{5}}\NR
x_{{1}}x_{{5}}&x_{{2}}x_{{5}}&-x_{{3}}x_{{5}}&-x_{{
4}}x_{{5}}&0\end {array} \right)
\]
with
$\Rank(\pi)=4 $.

The constants of motion are
\[
\begin{split}
H&=x_1+x_2+x_3+x_4+x_5\;\text{(Hamiltonian)},\\
F&=x_1x_3+x_1x_4+x_2x_4,\\
C&=x_2x_3x_5\;\text{(Casimir)}\,.
\end{split}
\]
These functions are obtained from the coefficients of the characteristic polynomial of $L$
\[
f(z)=\lambda^5- \left(x_1+x_2+x_3+x_4+x_5\right)\,  \lambda^3-2\sqrt{x_2x_3x_5}\, \lambda^2+\left(x_1x_3+x_1x_4+x_2x_4\right) \, \lambda \,.
\]
\end{example}

\section{Families of Lotka-Volterra systems} \label{families}

In the previous section we saw several examples of cubic systems  which (after  a simple change of variables)
are equivalent to  Lotka-Volterra systems. In this section we will describe all  subsets
$\Phi$ of the positive roots of  $A_n$ which  produce, after  a suitable change of variables, a
Lotka-Volterra system.

It can be verified that for a root system of type $A_n$  the only choice for the subset  $\Phi$  of $\Delta^+$  which transforms into a Lotka-Volterra system using the
substitution $x_i=2a_i^2$  is one of the following five.
\begin{enumerate}
\item $\Phi=\Pi,$
\item $\Phi=\Pi\cup\{\alpha_2+\alpha_3+\cdots+\alpha_{n-1}\},$
\item $\Phi=\Pi\cup\{\alpha_1+\alpha_2+\cdots+\alpha_{n-1} \},$
\item $\Phi=\Pi\cup\{\alpha_2+\alpha_3+\cdots+\alpha_n \},$
\item $\Phi=\Pi\cup\{\alpha_1+\alpha_2+\cdots+\alpha_n \}\,.$
\end{enumerate}

Case (1) gives rise to the KM system while case (5) gives rise to the periodic KM system.

Case (2) corresponds to the Lax equation $\dot{L}=[B,L]$ with $L$ matrix
\[L=\begin{pmatrix}
0      &a_1    &0      &\cdots &  0    &0      &0      &0      \\
a_1    &0      &a_2    &0      &       &  0    &a_{n+1}&0      \\
0      &a_2    &0      &a_3    &\ddots &       &  0    &0      \\
\vdots &0      &a_3    &\ddots &\ddots &       &       &0      \\
  0    &       &\ddots &\ddots &   0   &a_{n-2}&  0    &\vdots \\
  0    &   0   &       &       &a_{n-2}&   0   &a_{n-1}& 0     \\
  0    &a_{n+1}& 0     &       &  0    &a_{n-1}& 0     &a_n    \\
0      &0      & 0     & 0     & \cdots&   0   &a_n    &0
\end{pmatrix},
\]
The matrix $B$ is defined using the method described in section \Ref{Procedure} as
\[\begin{pmatrix}
0         &0       &a_1a_2 &0      &\cdots         &   0&0         &-a_1a_{n+1}    &0         \\
0         &0       &0      &a_2a_3 &               & 0   &-a_{n-1}a_{n+1}&0              &-a_na_{n+1}\\
-a_1a_2   & 0      &0      &0      &\ddots         &    &   0        &-a_2a_{n+1}&0          \\
0         &-a_2a_3 &0      &\ddots &         &     &          &     0          &0     \\
\vdots    &        &\ddots & &               &   \ddots   &\ddots    &               & \vdots     \\
0         &0       &       &       &         \ddots      &    \ddots   & 0        &a_{n-1}a_{n-2} &     0      \\
  0       & a_{n-1}a_{n+1} &0     &               & \ddots       &    0   &   0           &0       & a_{n-1}a_n \\
a_1a_{n+1}&0           &a_2a_{n+1}& 0    &\ddots  &-a_{n-1}a_{n-2}&      0      & 0             & 0            \\
0         &a_na_{n+1}& 0    & 0 & \cdots        &      0        &  -a_{n-1}a_n &0       &0
\end{pmatrix}.
\]

After substituting $x_i=2a_i^2$ for $i = 1,\dots,n+1$, the Lax pair $B,L$ becomes equivalent to the following equations of motion:
$$
\begin{array}{rcll}
\dot{x}_1&=&x_1(x_2-x_{n+1}),&\\
\dot{x}_2&=&x_2(x_3-x_1-x_{n+1}),&\\
\dot{x}_i&=&x_i(x_{i+1}-x_{i-1}),& i=3,4,\ldots ,n-2,n\\
\dot{x}_{n-1}&=&x_{n-1}(x_n-x_{n-2}+x_{n+1}),&\\
\dot{x}_{n+1}&=&x_{n+1}(x_1+x_2-x_{n-1}-x_n).&
\end{array}
$$
It is easily verified that for $n$ even, the rank of the Poisson matrix is $n$ and the function
	$f=x_2x_3\cdots x_{n-1}x_{n+1}$ is the Casimir of the system,
while for $n$ odd, the rank of the Poisson matrix is $n-1$ and the 	functions
	$f_1=x_1x_3\cdots x_n=\sqrt{\det L}$ and
	$f_2=x_2x_3\cdots x_{n-1}x_{n+1}$ are the Casimirs.

Case (3) corresponds to the Lax pair $L,B$ where the matrices $L,B$ are given by

\[L=\begin{pmatrix}
0      &a_1    &0      &\cdots &       &0      &a_{n+1}&0      \\
a_1    &0      &a_2    &0      &       &       &0      &0      \\
0      &a_2    &0      &a_3    &\ddots &       &       &0      \\
\vdots &0      &a_3    &\ddots &\ddots &       &       &\vdots \\
       &       &\ddots &\ddots &   0   &a_{n-2}&  0    &       \\
  0    &       &       &       &a_{n-2}&   0   &a_{n-1}& 0     \\
a_{n+1}& 0     &       &       &  0    &a_{n-1}& 0     &a_n    \\
0      &0      & 0     & \cdots&       &   0   &a_n    &0
\end{pmatrix},
\]

\[B=\begin{pmatrix}
0      &0      &a_1a_2 &0      &\cdots         &-a_{n-1}a_{n+1}&0              &-a_na_{n+1}\\
0      &0      &0      &a_2a_3 &               &      0        &-a_1a_{n+1}    &0          \\
-a_1a_2&0      &0      &0      &\ddots         &               &    0          &0          \\
0      &-a_2a_3&0      &\ddots &\ddots         &               &               &\vdots     \\
\vdots &       &\ddots &\ddots &               &               &               &           \\
0      &       &       &       &               &               &a_{n-1}a_{n-2} &     0      \\
  a_{n-1}a_{n} &   0   &       &               &               &   0           &0       & a_{n-1}a_n \\
0&a_{1}a_{n+1} &    0  &       &-a_{n-1}a_{n-2}&      0        & 0             & 0            \\
a_na_{n+1}     &0      & 0     & \cdots        &      0        &  -a_{n-1}a_n  &0       &0
\end{pmatrix}.
\]
After substituting $x_i=2a_i^2$ for $i=1,\dots, n+1$, the Lax pair $(B,L)$ becomes equivalent to the following equations of motion:
\begin{eqnarray*}
\dot{x}_1&=&x_1(x_2-x_{n+1})\\
\dot{x}_i&=&x_i(x_{i+1}-x_{i-1}), \ i=2,3,4,\ldots ,n-2,n\\
\dot{x}_{n-1}&=&x_{n-1}(x_n-x_{n-2}+x_{n+1})\\
\dot{x}_{n+1}&=&x_{n+1}(x_1-x_n-x_{n-1}).
\end{eqnarray*}
For $n$ even, the rank of the Poisson matrix is $n$ and
	the function $f=x_1x_2\cdots x_{n-1}x_{n+1}$ is the Casimir,
while for $n$ odd, the rank of the Poisson matrix is $n-1$ and the functions
	$f_1=x_1x_3x_5\cdots x_n=\sqrt{\det L}$ and
	$f_2=x_1x_2\cdots x_{n-1}x_{n+1}$ are Casimirs.

The system obtained in case (4) turns out to be isomorphic  to the one in case (3).  In fact, the change of variables $u_{n+1-i}=-x_i$ for $i=1,2,\ldots,n$ and $u_{n+1}=-x_{n+1}$ in case (3) gives the corresponding system of case (4).

\section{Two  Lax pair techniques} \label{2tech}

In this section we present two techniques that we use to prove the integrability of the generalized Lotka-Volterra systems.  The first one is due to  Deift, Li, Nanda and Tomei in \cite{DLNT}. It was used to establish the complete integrability of the full Kostant Toda lattice.  The traces of powers of $L$ were not enough to prove integrability, therefore the method of chopping was used to obtain additional integrals.
First we describe the method:
For $k=0, \dots , \left\lfloor { n-1\over 2}\right\rfloor$,\, denote by $( L-
\lambda \, { \rm Id})_{ (k)}$ the result of removing the first $k$
rows and last $k$ columns from $L- \lambda \,{\rm  Id}$, and let
\bd {\rm det} \ ( L- \lambda \, { \rm Id})_{ (k)} = E_{0k} \lambda
^{n- 2k} + \dots + E_{n-2k,k} \ . \ed

Set \bd {  {\rm det} \ ( L- \lambda \, { \rm Id})_{ (k)}  \over
E_{0k}} = \lambda^{ n-2k} + I_{1k} \lambda ^ {n-2k-1} + \dots +
I_{n-2k,k} \ . \ed The functions $I_{rk}$, $r=1, \dots, n-2k$, are
constants of motion for the FKT lattice.

\bigskip
\begin{example}  We consider in detail the $gl(3,{\bf C})$
case of the full Toda.  Let

\bd L= \begin{pmatrix} f_1&1&0\cr g_1&f_2&1 \cr h_1&g_2 &f_3
\end{pmatrix} \ ,
\ed  and take $B$ to be the strictly lower part of $L$. The function  $H_2={1 \over 2}  \tr L^2$ is  the Hamiltonian,   and using  a suitable  linear  Poisson bracket  the equations  \bd \dot x=\{H_2, x\} \ed are equivalent to \bd
\begin{array}{lcl}
\dot f_1 &=& -g_1 \cr \dot f_2 &=& g_1-g_2 \cr \dot f_3 &=& g_2
\cr \dot g_1 &=&g_1(f_{1}-f_2) -h_1\cr \dot g_2 &=&g_2(f_{2}-f_3)
+h_1\cr \dot h_1 &=& h_1 (f_{1}-f_3) \ .
\end{array}
\ed
Note that $H_1=f_1+f_2+f_3$ while $H_2= {1 \over 2}
(f_1^2+f_2^2+f_3^2)+g_1+g_2$.

The chopped matrix is given by \bd
\begin{pmatrix}
g_1& f_2-\lambda \cr h_1 & g_2
\end{pmatrix} \ .
\ed

The determinant of this matrix is $h_1 \lambda +g_1 g_2-h_1 f_2$
and one  obtains the rational integral \be I_{11}={ g_1 g_2 -h_1 f_2
\over h_1} \ . \label{a8}  \ee

Note that the phase space is six dimensional, we have two Casimirs
$(H_1, I_{11})$ and the functions $(H_2, H_3)$ are enough to
ensure integrability.

\end{example}

In the next example we use this technique to obtain the Casimir of a generalized Lotka-Volterra system.
\begin{example}
\noindent
 Consider the generalized Lotka-Volterra system defined by the Lax matrix
$$ L=
\begin{pmatrix}
0 &a_1&0&a_5&0\\
a_1 &0&a_2&0&0\\
0 &a_2&0&a_3&0\\
a_5 &0&a_3&0&a_4\\
0&0&0&a_4&0
\end{pmatrix}
$$
which  corresponds to the subset $\Phi=\{\alpha_1,\alpha_2,\alpha_3,\alpha_4,\alpha_1+\alpha_2+\alpha_3\}$.
According to section \ref{liealgebras} a suitable  choice of signs for the entries of $B$ gives rise to a Lotka-Volterra system.
However, there is  second choice of sings which results in  a different system.  Define  the matrix $B$ to be
$$
\begin{pmatrix}
0& 0 &a_1a_2+a_3a_5&0&a_4a_5\\
0 &0&0&a_2a_3+a_1a_5&0\\
-a_1a_2-a_3a_5&0 &0&0&a_3a_4\\
0 &-a_2a_3-a_1a_5&0&0&0\\
-a_4a_5 &0&-a_3a_4&0&0\\
\end{pmatrix} \ .
$$
In this case  the Lax equation $\dot{L}=[B,L]$ corresponds to the following system
\begin{align*}
\dot{a_1}&=a_1a_2^2+a_1a_5^2+2a_2a_3a_5\\
\dot{a_2}&=a_2a_3^2-a_2a_1^2\\
\dot{a_3}&=a_3a_4^2-a_3a_2^2-a_3a_5^2-2a_1a_2a_5\\
\dot{a_4}&=-a_4a_5^2-a_4a_3^2\\
\dot{a_5}&=-a_5a_1^2+a_5a_3^2+a_5a_4^2  \ .
\end{align*}

The Hamiltonian of the system is $H=\dfrac{1}{2}\left(a_1^2+a_2^2+a_3^2+a_4^2+a_5^2\right)$ and the Poisson matrix (of  rank $4$) is
$$
\begin{pmatrix}
0&a_1a_2&2a_2a_5&0&a_1a_5\\
-a_1a_2&0&a_2a_3&0&0\\
-2a_2a_5&-a_2a_3&0&a_3a_4&-a_3a_5\\
0&0&-a_3a_4&0&-a_4a_5\\
-a_1a_5&0&a_3a_5&a_4a_5&0
\end{pmatrix} \ .
$$
The system is integrable with constants of motion $H=\dfrac{1}{2}\left(a_1^2+a_2^2+a_3^2+a_4^2+a_5^2\right)$ and
$$
F=\tr\left(\frac{L^4}{4}\right)=
\frac{1}{2}a_1^4+a_1^2a_5^2+\frac{1}{2}a_5^4+a_1^2a_2^2+2a_1a_5a_2a_3+a_3^2a_5^2+a_4^2a_5^2+\frac{1}{2}a_2^4+a_2^2a_3^2+\frac{1}{2}a_3^4+a_4^2a_3^2+\frac{1}{2}a_4^4.
$$
The Casimir of the system is
$
C=a_2^2-\dfrac{a_1a_2a_3}{a_5}
$
and  may be  obtained by the method  chopping as follows.
We have
$$
x\cdot I_5-L=
\begin{pmatrix}
x &-a_1&0&-a_5&0\\
-a_1 &x&-a_2&0&0\\
0 &-a_2&x&-a_3&0\\
-a_5 &0&-a_3&x&-a_4\\
0&0&0&-a_4&x
\end{pmatrix}
$$
and the one-chopped matrix is
$$
\begin{pmatrix}
-a_1 &x&-a_2&0\\
0 &-a_2&x&-a_3\\
-a_5 &0&-a_3&x\\
0&0&0&-a_4
\end{pmatrix}
$$
with determinant
$
a_4a_5x^2+a_1a_2a_3a_4-a_2^2a_4a_5.
$
Dividing the constant term of this polynomial  by the leading term  $a_4a_5$ we obtain the Casimir $C$.

\end{example}

The second method that we use is an old  recipe of Moser.
Moser in \cite{moser} describes a relation between the KM system   and the non--periodic Toda lattice. The procedure is
the following: Form $L^2$  which is not anymore a tridiagonal matrix but is similar to one.  Let $\{e_1, e_2, \dots, e_n \}$ be the standard
basis of ${\bf R}^n$, and $E_o= \{ {\rm span}\, e_{2i-1}, \,  i=1,2, \dots \}$, $E_e= \{ {\rm span}\, e_{2i}, \, i=1,2, \dots \}$. Then $L^2$ leaves
$E_o$,  $E_e$ invariant and reduces in  each of these spaces to a tridiagonal symmetric Jacobi matrix.
 For example, if we omit  all even columns and all even  rows we
obtain a tridiagonal Jacobi matrix and the entries of this new matrix  define the transformation from the KM--system
to the Toda lattice. We illustrate with a simple example where $n=5$.

We use the symmetric version of the KM system Lax pair   given by
\bd
L=\begin{pmatrix} 
	0   & a_1 & 0   &  0  &  0  \cr
	a_1 & 0   & a_2 &  0  &  0  \cr
	0   & a_2 & 0   & a_3 &  0  \cr
	0   & 0   & a_3 &  0  & a_4 \cr
	0   & 0   & 0   & a_4 &  0
\end{pmatrix}  \ .
\ed

It is simple to calculate that  $L^2$ is the matrix

\bd
\begin{pmatrix} 
a_1^2   &     0       & a_1 a_2     &     0       &    0    \cr
0       & a_1^2+a_2^2 &     0       &   a_2 a_3   &    0    \cr
a_1 a_2 &     0       & a_2^2+a_3^2 &      0      & a_3 a_4 \cr
0       & a_2 a_3     &    0        & a_3^2+a_4^2 &    0    \cr
0       &     0       & a_3 a_4     &      0      &   a_4^2
\end{pmatrix} \ .
\ed
Omitting even  columns and even rows of $L^2$ we obtain the matrix
\bd
\begin{pmatrix}  
a_1^2   &    a_1 a_2  &    0    \cr
a_1 a_2 & a_2^2+a_3^2 & a_3 a_4 \cr
0       &   a_3 a_4   &  a_4^2
\end{pmatrix} \ .
\ed
This is a tridiagonal Jacobi matrix. It is natural to define   new variables $A_1=a_1 a_2$, $A_2=a_3 a_4$, $B_1=a_1^2$, $B_2=a_2^2+a_3^2$, $B_3=a_4^2$. The new
 variables $A_1,A_2, B_1,B_2, B_3$ satisfy the Toda lattice  equations.

This procedure shows that the KM-system  and the Toda lattice are closely related: The explicit  transformation
 which is  due to H\'enon
maps one system to the other. The mapping in the general case  is given by
\be
A_i=-{ 1 \over 2} \sqrt {a_{2i} a_{2i-1}} \  , \qquad  B_i= { 1 \over 2}\left( a_{2i-1}+a_{2i-2} \right)  \label{a25} \ .
\ee
The equations satisfied by the new variables $A_i$, $B_i$ are given by:

\begin{displaymath}
\begin{array}{lcl}
 \dot A _i& = & A_i \,  (B_{i+1} -B_i )    \\
   \dot B _i &= & 2 \, ( A_i^2 - A_{i-1}^2 ) \ .
\end{array}
\end{displaymath}
These are precisely the Toda equations in Flaschka's form.

This idea of Moser was applied with success to establish transformations   from the generalized Volterra  lattices of Bogoyavlensky \cite{bog1, bog2}  to generalized Toda systems.
The relation between the Volterra systems of type $B_n$ and $C_n$ and the corresponding Toda systems is in \cite{damianou02}.  The similar construction of  the Volterra lattice   of type $D_n$ and the generalized
Toda lattice of type $D_n$   is in \cite{damianou04}. We use this method in the next section to obtain a missing integral for some   generalized Lotka-Volterra systems.

\section{$2$-diagonal  systems} \label{2diag}
We define a family of systems with a cubic Hamiltonian vector field. We present each such system in Lax pair form
$\dot{L}=[B,L]$ which allows us to obtain a large family of first integrals, $H_i=\tr(L^i)$. Additional integrals are obtained by the method of Moser  discribed in the previous section. In the examples we present, these integrals are enough to ensure
the Liouville integrability of the systems. We believe that all these systems are Liouville integrable.

We begin with the definition of the matrices $L$ and $B$.
For convenience we let $d_i$ denote the $i^{th}$ diagonal starting from the upper right corner and moving towards the main diagonal.
 We take $L$ to be an $n\times n$ symmetric matrix with the only  non-zero entries on two diagonals $d_m$ and $d_{n-1}$  where $n \geqslant 2m$ and $m \geqslant 2$. Note that for $m=1$ we obtain the periodic KM system.

The matrix $L$ is given by
\[
L=\begin{pmatrix}
0      &a_1    &0      &\cdots   &0      &  a_n  &0      &\cdots &0        \\
a_1    &0      &a_2    &0        &       &  0    &a_{n+1}&\ddots &\vdots   \\
0      &a_2    &0      &a_3      &\ddots &       &\ddots &\ddots &0        \\
\vdots &0      &a_3    &\ddots   &\ddots &       &       &0      &a_{n+m-1}\\
  0    &       &       &\ddots   &       &       &       &       &      0  \\
  a_n  &   0   &       &         &       &       &a_{n-2}&  0    &\vdots   \\
  0    &a_{n+1}&\ddots &         &       &a_{n-2}&   0   &a_{n-2}& 0       \\
 \vdots&\ddots & \ddots&0        &       &  0    &a_{n-2}& 0     &a_{n-1}  \\
0      &\cdots & 0     &a_{n+m-1}& 0     & \cdots&   0   &a_{n-1}&0
\end{pmatrix}.
\]
That is, $L$ is a symmetric $n \times n$ matrix whose non-zero upper diagonals are:
\[\displaystyle
\begin{array}{ccl}
	d_{n-1}	&=& (a_{1},a_{2}, \dots, a_{n-1})\NR
	d_m 	&=& (a_{n}, a_{n+1},\dots, a_{n+m-1})
\end{array}  \]
To put it in the terminology of section \Ref{Procedure} this matrix has variables in the
positions corresponding to the simple roots and the positive roots of length $n-m$, i.e.
$$L=\displaystyle \sum_{\alpha_i\in\Phi}a_i(X_{\alpha_i}+X_{-\alpha_i}),$$
where
$$
\Phi=\{\alpha_1,\alpha_2,\ldots,\alpha_{n-1},\alpha_1+\alpha_2+\ldots+\alpha_{n-m},
\ldots,\alpha_{m}+\alpha_{m+1}+\ldots+\alpha_{n-1}\}.
$$
By considering the set
$
\Psi =
\left\{
\alpha+\beta \ | \ \alpha,  \beta \in \Phi \cup \Phi^{-},  \alpha+\beta \in \Delta^{+}
\right\}
$
we define $B$ to be the matrix
\begin{equation}
\label{B_Matrix}
B=
\sum c_{ij} a_i a_j \left(\left[X_{\alpha_i},X_{\alpha_j}\right]+\left[X_{-\alpha_i},X_{-\alpha_j}\right]\right),
\end{equation}
where the non-zero terms are  taken over all
$\alpha_i+\alpha_j \in \Psi$
with
$\alpha_i,\alpha_j\in\Phi\cup\Phi^-$
and
$c_{ij}=\pm 1$.
We compute the signs $c_{ij}$ in a way that leads to a consistent Lax pair. It turns out that $B$ is the $n \times n$ skew-symmetric matrix with non-zero upper diagonals:
\begin{equation}\label{B_Matrix_diagonals}
\begin{array}{{r@{\hspace{3pt}}c@{\hspace{3pt}}l@{\hspace{3pt}}}}
	d_{n-2}	&=& (a_{1}a_{2}, a_{2}a_{3},\dots , a_{n-2}a_{n-1}),\\
	d_{m+1}	&=& (-a_{n-m}a_{n}, -a_{n-m+1}a_{n+1}-a_{1}a_{n},\dots ,-a_{n-1}a_{n+m-1}-a_{m}a_{n+m-2}, -a_{m}a_{n+m-1}),\\
	d_{m-1}	&=& (a_{n-m+1}a_{n}+a_{1}a_{n+1}, a_{n-m+2}a_{n+1}+a_{2}a_{n+2},\dots, a_{n-1}a_{n+m-2}+a_{m-1}a_{n+m-1} ).
\end{array}
\end{equation}

The poisson bracket $\{\,,\}$ is determined by the $N \times N$ Poisson matrix $\Gp=q-q^t$, where $N=n+m-1$, and the non-zero entries of $q$ are given by:
\begin{equation}\label{Poisson_Matrix}
\displaystyle
\begin{array}{lcll}
	q_{i,i+n}			&=& a_{i}a_{i+n}			&\text{for } 1 \leqslant i \leqslant m-1,\NR
	q_{i,i+n-1}			&=& -a_{i}a_{i+n-1}			&\text{for } 1 \leqslant i \leqslant m,\NR
	q_{i+n-m-1,i+n-1}	&=& a_{i+n-1}a_{i+n-m-1}	&\text{for } 1 \leqslant i \leqslant m,\NR
	q_{i+n-m,i+n-1}		&=& -a_{i+n-1}a_{i+n-m}		&\text{for } 1 \leqslant i \leqslant m-1,\NR
	q_{i,i+1}			&=& a_{i}a_{i+1}			&\text{for } 1 \leqslant i \leqslant n-2,\NR
	q_{i+n-1,i+n}		&=& 2a_{i}a_{i+n-m}			&\text{for } 1 \leqslant i \leqslant m-1\,.
\end{array}
\end{equation}

\subsection{Example.}
We illustrate in detail the results with a specific example when $m=4$, $n=10$ and $N=13$.
Here $L$ is a $10\times10$ matrix with two diagonals $d_4$ and $d_9$, i.e.
\[
\displaystyle 
L= 
\begin{pmatrix}
0&a_{{1}}&0&0&0&0&a_{{10}}&0&0&0\\
a_{{1}}&0&a_{{2}}&0&0&0&0&a_{{11}}&0&0\\
0&a_{{2}}&0&a_{{3}}&0&0&0&0&a_{{12}}&0\\
0&0&a_{{3}}&0&a_{{4}}&0&0&0&0&a_{{13}}\\
0&0&0&a_{{4}}&0&a_{{5}}&0&0&0&0\\
0&0&0&0&a_{{5}}&0&a_{{6}}&0&0&0\\
a_{{10}}&0&0&0&0&a_{{6}}&0&a_{{7}}&0&0\\
0&a_{{11}}&0&0&0&0&a_{{7}}&0&a_{{8}}&0\\
0&0&a_{{12}}&0&0&0&0&a_{{8}}&0&a_{{9}}\\
0&0&0&a_{{13}}&0&0&0&0&a_{{9}}&0
\end{pmatrix}.
\]
The matrix $B $ is defined by equation \Ref{B_Matrix}
and the Poisson bracket $\{\,,\}$ is determined by equations \Ref{Poisson_Matrix}.
The resulting Lax pair, $(L,B)$, is equivalent to the following equations of motion:
\begin{eqnarray*}
\dot{a _{1}}  &=& a_{1}a_{2}^{2}+a_{1}a_{11}^{2}-a_{1}a_{10}^{2}, \NR
\dot{a _{2}}  &=& a_{2}a_{3}^{2}-a_{1}^{2}a_{2}+a_{2}a_{12}^{2}-a_{2}a_{11}^{2}, \NR
\dot{a _{3}}  &=& a_{3}a_{4}^{2}-a_{2}^{2}a_{3}+a_{3}a_{13}^{2}-a_{3}a_{12}^{2}, \NR
\dot{a _{4}}  &=& a_{4}a_{5}^{2}-a_{3}^{2}a_{4}-a_{4}a_{13}^{2}, \NR
\dot{a _{5}}  &=& a_{5}a_{6}^{2}-a_{4}^{2}a_{5}, \NR
\dot{a _{6}}  &=& a_{6}a_{7}^{2}-a_{5}^{2}a_{6}+a_{6}a_{10}^{2}, \NR
\dot{a _{7}}  &=& a_{7}a_{8}^{2}-a_{6}^{2}a_{7}+a_{7}a_{11}^{2}-a_{7}a_{10}^{2}, \NR
\dot{a _{8}}  &=& a_{8}a_{9}^{2}-a_{7}^{2}a_{8}+a_{8}a_{12}^{2}-a_{8}a_{11}^{2}, \NR
\dot{a _{9}}  &=& -a_{8}^{2}a_{9}+a_{9}a_{13}^{2}-a_{9}a_{12}^{2}, \NR
\dot{a} _{10} &=& a_{7}^{2}a_{10}-a_{6}^{2}a_{10}+a_{1}^{2}a_{10}+2a_{1}a_{7}a_{11}, \NR
\dot{a} _{11} &=& a_8^2a_{11}-a_7^2a_{11}+a_2^2a_{11}-a_1^2a_{11}+2a_2a_8a_{12}-2a_1a_7a_{10}, \NR
\dot{a} _{12} &=& a_9^2a_{12}-a_8^2a_{12}+a_3^2a_{12}-a_2^2a_{12}+2a_3a_9a_{13}-2a_2a_8a_{11}, \NR
\dot{a} _{13} &=& -a_{9}^{2}a_{13}+a_{4}^{2}a_{13}-a_{3}^{2}a_{13}-2a_{3}a_{9}a_{12}. \NR
\end{eqnarray*}
The Hamiltonian of the system is $H_2 = \frac{1}{2}\left(a_1^2 + a_2^2 + \cdots + a_{13}^2\right)$ and the Poisson matrix has rank $12$.
The following constants of motion
$$H_i=\tr L^i, \ \ i=2,4,6,7,8,9$$
together with the Casimir, $C=\det L$, ensure the integrability of the system.
\subsection{Special case with two diagonals, $m=2$}
In this subsection we consider the case where $m=2$.
The matrix $L$ is defined by

\[L=\left(
\begin {array}{ccccccc}
0     & a_1   & 0    &\cdots&  0    &  a_n  & 0       \LR
a_1   & 0     & a_2  &\ddots&       &  0    &a_{n+1}  \LR
0     & a_2   &  0   &\ddots&       &       &0        \LR
\vdots&\ddots &\ddots&\ddots&       &   0   &\vdots   \LR
 0    &       &      &      &       &a_{n-2}&0        \LR
a_n   &  0    &      &  0   &a_{n-2}&  0    &a_{n-1}  \LR
0     &a_{n+1}&  0   &\cdots&   0   &a_{n-1}&0
\end {array} \right)
\]
and corresponds to the subset $\Phi$ of the positive roots containing the simple roots and the roots of length $n-2$.
The matrix $B$ is defined by equation \Ref{B_Matrix} and its upper triangular part is
\[\begin{pmatrix}
0      &0      &a_1a_2 &0              &\cdots&   0     & -a_{n-2}a_{n} &0     &a_1a_{n+1}+a_{n-1}a_n \\
0      &0      &0      &a_2a_3         &      &   \;    &      0        &-a_1a_{n}-a_{n-1}a_{n+1}&0   \\
\vdots & \ddots&0      &0              &\ddots&   \;    &               &    0                   &-a_2a_{n+1}\\
       &       &       &\ddots         &\ddots&   \;    &               &                        &     0     \\
       &       &       &               &      &   \;    &   \ddots      &    0                   &    \vdots \\
       &       &       &               &      &   \;    &     0         &a_{n-3}a_{n-2}          &     0     \\
       &       &       &               &      &   \;    &     0         &0                       & a_{n-2}a_{n-1}\\
\vdots &       &       &               &      &         &   \ddots      & 0                      &0              \\
0      &\cdots &       &               &      &         &   \cdots      & 0                      &0    \end{pmatrix},
\]
The Lax equation $\dot{L} =[B,L]$ is equivalent to the following system:
\begin{eqnarray*}
	\dot{a}_1     &=&	a_1a^2_2+a_1a^2_{n+1}-a_1a^2_n, \NR
	\dot{a}_2     &=&	a_2a^2_3-a^2_1a_2-a_2a^2_{n+1}, \\
	\vdots  \     & &	\quad\quad\quad\quad	\vdots  \\
	\dot{a}_i     &=&	a_ia^2_{i+1}-a^2_{i-1}a_i, \quad \quad \quad i = 3, 4, \dots, n-3\\
	\vdots \      & &	\quad\quad\quad\quad	\vdots  \\
	\dot{a}_{n-2} &=& 	a_{n-2}a^2_n-a^2_{n-3}a_{n-2}+a_{n-2}a^2_{n-1},\NR
	\dot{a}_{n-1} &=& 	a_{n-1}a^2_{n+1}-a^2_{n-2}a_{n-1}-a_{n-1}a^2_n,\NR
	\dot{a}_n     &=& 	a^2_1a_n+a^2_{n-1}a_n-a^2_{n-2}a_n+2a_{1}a_{n-1}a_{n+1},\NR
	\dot{a}_{n+1} &=&   a^2_2a_{n+1}-a^2_1a_{n+1}-a_{n+1}a^2_{n-1}-2a_1a_{n-1}a_n\,.
\end{eqnarray*}

\noindent
The Poisson matrix $\Gp$ is defined by equations \Ref{Poisson_Matrix} and its upper triangular part is
\[
\begin{pmatrix}
0      &a_1a_2 &    0  & \cdots        &      &       & 0             & -a_1a_n      & a_1a_{n+1}     \\
\vdots &0      &a_2a_3 &     0         &      & \;    &               & 0            & -a_2a_{n+1}    \\
       &       &0      & a_3a_4        &      & \;    &               &              &      0         \\
       &       &       & \ddots        &      & \ddots& 0             &              &                \\
       &       &       &               &\ddots& \ddots& 0             & 0            &    \vdots      \\
       &       &       &               &      & 0     & a_{n-2}a_{n-1}& a_{n-2}a_{n} &     0          \\
       &       &       &               &      & \;    & 0             & -a_{n-1}a_n  & a_{n-1}a_{n+1} \\
\vdots &       &       &               &      & \;    &               & 0            & 2a_1a_{n-1}    \\
0      &\cdots &       &               &      & \;    &               & \cdots       & 0
\end{pmatrix},
\]
For $n=9$ the corresponding system is
\begin{eqnarray*}
	\dot{a}_1    &=&    a_1a^2_2+a_1a^2_{10}-a_1a^2_9,\NR
	\dot{a}_2    &=&    a_2a^2_3-a^2_1a_2-a_2a^2_{10},\NR
	\dot{a}_3    &=&    a_3a^2_4-a^2_2a_3,\NR
	\dot{a}_4    &=&    a_4a^2_5-a^2_3a_4,\NR
	\dot{a}_5    &=&    a_5a^2_6-a^2_4a_5,\NR
	\dot{a}_6    &=&    a_6a^2_7-a^2_5a_6,\NR
	\dot{a}_7    &=&    a_7a^2_8-a^2_6a_7+a_7a^2_9,\NR
	\dot{a}_8    &=&    -a^2_7a_8+a_8a^2_{10}-a_8a^2_9,\NR
	\dot{a}_9    &=&    a^2_1a_9+a^2_8a_9-a^2_7a_9+2a_1a_8a_{10},\NR
	\dot{a}_{10} &=&    a^2_2a_{10}-a^2_1a_{10}-a^2_8a_{10}-2a_1a_8a_9.
\end{eqnarray*}

It has Lax representation $\dot{L}=[B,L]$ with
\[
L=
\begin {pmatrix}
0&a_{{1}}&0&0&0&0&0&a_{{9}}&0        \NR
a_{{1}}&0&a_{{2}}&0&0&0&0&0&a_{{10}} \NR
0&a_{{2}}&0&a_{{3}}&0&0&0&0&0        \NR
0&0&a_{{3}}&0&a_{{4}}&0&0&0&0        \NR
0&0&0&a_{{4}}&0&a_{{5}}&0&0&0        \NR
0&0&0&0&a_{{5}}&0&a_{{6}}&0&0        \NR
0&0&0&0&0&a_{{6}}&0&a_{{7}}&0        \NR
a_{{9}}&0&0&0&0&0&a_{{7}}&0&a_{{8}}  \NR
0&a_{{10}}&0&0&0&0&0&a_{{8}}&0
\end{pmatrix}
\]
and the matrix $B$ is defined by relation \Ref{B_Matrix} and given by equations \Ref{B_Matrix_diagonals}.

We conjecture that for any $n$ the corresponding system is integrable.
For $n$ even, the system has $n+1$ variables and the Poisson matrix has rank $n$ and thus the Poisson structure has one Casimir.
The traces of $L$ give $n/2$ functionally independent first integrals in involution. Hence the system is integrable
in the sense of Liouville. The Casimir is $$C=\det{L}=(a_3a_5\ldots a_{n-5}a_{n-3}a_na_{n+1}-a_1a_3\ldots a_{n-3}a_{n-1})^2.$$

For $n$ odd, the system has $n+1$ variables and the Poisson matrix has rank $n+1$. Therefore the
Poisson structure is non-degenerate with no Casimirs. The traces $\tr(L^i)$ give only $\frac{n+1}{2}-1$
functionally independent first integrals in involution.
For the integrability of the system we need one more
constant of motion which we obtain using a procedure due to Moser described in section \Ref{2tech}.

We give two examples for $n=7,n=9$.

\begin{example}
Consider the following matrices
\[
L=
\begin{pmatrix}
0&a_{{1}}&0&0&0&a_{{7}}&0\NR
a_{{1}}&0&a_{{2}}&0&0&0&a_{{8}}\NR
0&a_{{2}}&0&a_{{3}}&0&0&0\NR
0&0&a_{{3}}&0&a_{{4}}&0&0\NR
0&0&0&a_{{4}}&0&a_{{5}}&0\NR
a_{{7}}&0&0&0&a_{{5}}&0&a_{{6}}\NR
0&a_{{8}}&0&0&0&a_{{6}}&0
\end{pmatrix},
\]
\[
\La_o(L^2)=
\begin{pmatrix}
a_1^2+a_2^2+a_8^2 & a_2a_3      & a_1a_7+a_6a_8 \NR
a_2a_3            & a_3^2+a_4^2 & a_4a_5        \NR
a_1a_7+a_6a_8     & a_4a_5      & a_5^2+a_6^2+a_7^2
\end{pmatrix}
\]
We define a new set of variables $A_1=a_2a_3,A_2=a_4a_5,A_3=a_1a_7+a_6a_8,B_1=a_1^2+a_2^2+a_8^2,B_2=a_3^2+a_4^2\text{ and } B_3=a_5^2+a_6^2+a_7^2$.
These variables satisfy  the periodic Toda   equations which are equivalent to the Lax equation $\dot{\La}_o(L^2)=[C,\La_o(L^2)]$ where
\[
\La_o(L^2)=
\begin{pmatrix}
B_{{1}}&A_{{1}}&A_{{3}}\NR
A_{{1}}&B_{{2}}&A_{{2}}\NR
A_{{3}}&A_{{2}}&B_{{3}}
\end{pmatrix}
\]
and
\[
C=
\begin{pmatrix}
 0       &  A_{{1}} &-A_{{3}}\NR
-A_{{1}} &    0     & A_{{2}}\NR
 A_{{3}} & -A_{{2}} &    0
\end{pmatrix}.
\]

This system has two Casimirs $B_1+B_2+B_3$ and $A_1A_2A_3$. The Casimir $B_1+B_2+B_3$ expressed as a function of the original variables gives the Hamiltonian while the Casimir $A_1A_2A_3$  gives the extra integral
\[
A_1A_2A_3=a_2a_3a_4a_5 \left( a_1a_7+a_6a_8\right).
\]

We could also obtain this integral from the system $\dot{\La}_e(L^2)=\left[C,\La_e(L^2)\right]$ where

\[
\La_e(L^2)=
\begin{pmatrix}
a_1^2+a_7^2   & a_1a_2          & a_5a_7      & a_1a_8 + a_6a_7 \NR
a_1a_2        & a_2^2+a_3^2     & a_3a_4      & a_2a_8          \NR
a_5a_7        & a_3a_4          & a_4^2+a_5^2 & a_5a_6          \NR
a_1a_8+a_7a_6 & a_2a_8          & a_5a_6      & a_6^2+a_8^2
\end{pmatrix}
=
\begin{pmatrix}
B_{{1}}&A_{{1}}&A_{{4}}&A_{{6}}\NR
A_{{1}}&B_{{2}}&A_{{2}}&A_{{5}}\NR
A_{{4}}&A_{{2}}&B_{{3}}&A_{{3}}\NR
A_{{6}}&A_{{5}}&A_{{3}}&B_{{4}}
\end{pmatrix}
\]
and
\[
C=
\begin{pmatrix}
  0      &  A_{{1}} &  -A_{{4}} &  A_{{6}}\NR
-A_{{1}} &    0     &  A_{{2}}  & -A_{{5}}\NR
 A_{{4}} & -A_{{2}} &     0     &  A_{{3}}\NR
-A_{{6}} &  A_{{5}} &  -A_{{3}} &     0
\end{pmatrix}.
\]

This system is not the  full symmetric Toda lattice of  Deift, Li, Nanda and Tomei \cite{DLNT}.  Although the $L$ matrix is the same, the $C$ matrix is different.  This system has two polynomial Casimirs,
$B_1+B_2+B_3+B_4$ and $A_{{1}}A_{{2}}A_{{4}}+A_{{2}}A_{{3}}A_{{5}}$, with
\[
A_{{1}}A_{{2}}A_{{4}}+A_{{2}}A_{{3}}A_{{5}}=a_{{2}}a_{{3}}a_{{4}}a_{{5}} \left( a_{{1}}a_{{7}}+a_{{6}}a_{{8}} \right).
\]
\end{example}

\begin{example}
We take $L$ to be
\[
\begin{pmatrix}
0&a_{{1}}&0&0&0&0&0&a_{{9}}&0\NR
a_{{1}}&0&a_{{2}}&0&0&0&0&0&a_{{10}}\NR
0&a_{{2}}&0&a_{{3}}&0&0&0&0&0\NR
0&0&a_{{3}}&0&a_{{4}}&0&0&0&0\NR
0&0&0&a_{{4}}&0&a_{{5}}&0&0&0\NR
0&0&0&0&a_{{5}}&0&a_{{6}}&0&0\NR
0&0&0&0&0&a_{{6}}&0&a_{{7}}&0\NR
a_{{9}}&0&0&0&0&0&a_{{7}}&0&a_{{8}}\NR
0&a_{{10}}&0&0&0&0&0&a_{{8}}&0
\end{pmatrix}.
\]
The matrix
\[
\La_o(L^2)=
\begin{pmatrix}
a_1^2+a_2^2+a_{10}^2&a_2a_3&0&a_1a_9+a_8a_{10}\NR
a_2a_3&a_3^2+a_4^2&a_4a_5&0\NR
0&a_4a_5&a_5^2+a_6^2&a_6a_7\NR
a_1a_9+a_8a_{10}&0&a_6a_7&a_7^2+a_8^2+a_9^2
\end{pmatrix}
=
\begin{pmatrix}
B_{{1}}&A_{{1}}& 0&A_{{4}}\NR
A_{{1}}&B_{{2}}&A_{{2}}&0\NR
0&A_{{2}}&B_{{3}}&A_{{3}}\NR
A_{{4}}& 0&A_{{3}}&B_{{4}}
\end{pmatrix}
\]
produces the periodic-Toda lattice which can be written in Lax pair form 
$\dot{\La}_o(L^2)=\left[C,\La_o(L^2)\right]$ with
\[
C=
\begin{pmatrix} 
   0      &  A_{{1}} &     0     & -A_{{4}} \NR
-A_{{1}}  &      0   &   A_{{2}} &    0     \NR
   0      & -A_{{2}} &     0     &  A_{{3}} \NR
 A_{{4}}  &    0     &  -A_{{3}} &   0
\end{pmatrix}.
\]
This system also has two polynomial  Casimirs $B_1+B_2+B_3+B_4$ and $A_1A_2A_3A_4$. By writing the latter one in the original variables
we obtain the extra integral, namely
\[
A_1A_2A_3A_4=a_2a_3a_4a_5a_6a_7 \left( a_1a_9+a_{10}a_8 \right).
\]

The {intermediate} Toda system $\dot{\La}_e(L^2)=\left[C,\La_e(L^2)\right]$ with
\[\begin{split}
\La_e(L^2)&=
\begin{pmatrix}
a_1^2+a_9^2&a_1a_2&0&a_7a_9&a_1a_{10}+a_8a_9\NR
a_1a_2&a_2^2+a_3^2&a_3a_4&0&a_2a_{10}\NR
0&a_3a_4&a_4^2+a_5^2&a_5a_6&0\NR
a_7a_9&0&a_5a_6&a_6^2+a_7^2&a_7a_8\NR
a_1a_{10}+a_8a_9&a_2a_{10}&0&a_7a_8&a_8^2+a_{10}^2
\end{pmatrix}
\\
&=
\begin{pmatrix}
B_1&A_1&0&A_5&A_7\NR
A_1&B_2&A_2&0&A_6\NR
0&A_2&B_3&A_3&0\NR
A_5&0&A_3&B_4&A_4\NR
A_7&A_6&0&A_4&B_5
\end{pmatrix}
\end{split}
\]
and
\[
C=
\begin{pmatrix}
 0   & A_1 &  0  &-A_5 & A_7 \NR
-A_1 & 0   & A_2 &  0  &-A_6 \NR
0    &-A_2 &  0  & A_3 &  0  \NR
 A_5 &  0  &-A_3 &  0  & A_4 \NR
-A_7 & A_6 &  0  &-A_4 &  0
\end{pmatrix}.
\]
has two Casimirs
$B_1+B_2+B_3+B_4+B_5$ and $A_1A_2A_3A_5+A_2A_3A_4A_6$ and
\[
A_1A_2A_3A_5+A_2A_3A_4A_6=a_2a_3a_4a_5a_6a_7\left(a_1a_9+a_8a_{10} \right).
\]

Note that this intermediate Toda system is not of the type considered in \cite{damianou11}.

\end{example}

\subsection{Special case with two diagonals, $m=3$}

In this subsection we consider the case where $m=3$.

The matrix $L$ is given by
\[
L=
\begin {pmatrix}
0     & a_1   & 0     &\ldots& 0     &  a_n  &  0    & 0       \LR
a_1   & 0     & a_2   &\ddots&       &  0    &a_{n+1}&0        \LR
0     & a_2   &  0    &\ddots&       &       &0      &a_{n+2}  \LR
\vdots&\ddots &\ddots &\ddots&       &       &       &      0  \LR
0     &\ddots &       &      &       &\ddots & \ddots& \vdots  \LR
a_n   & 0     &       &      & \ddots&\ddots &a_{n-2}&0        \LR
0     &a_{n+1}&  0    &      & \ddots&a_{n-2}&  0    &a_{n-1}  \LR
0     &0      &a_{n+2}&  0   &\cdots &   0   &a_{n-1}&0
\end {pmatrix}
\]
That is $L$ is a symmetric $n \times n$ matrix whose non-zero upper diagonals are:
\[\displaystyle
\begin{array}{ccl}
	d_{n-1}	&=& (a_{1},a_{2},a_{3},a_{4},\dots, a_{n-1}),\NR
	d_3 	&=& (a_{n}, a_{n+1}, a_{n+2}).
\end{array}  \]
It corresponds to the subset $\Phi$ of the positive roots of the root system of type $A_{n-1}$ containing the simple roots
and the roots of length $n-3$. The matrix $B$ constructed using the procedure described in \Ref{B_Matrix} is an $n \times n$
skew-symmetric matrix whose non-zero upper diagonals are:
\[\displaystyle
\begin{array}{ccl}
	d_{n-2}	&=& (a_{1}a_{2}, a_{2}a_{3}, a_{3}a_{4},\dots, a_{n-2}a_{n-1}),\NR
	d_4 		&=& (-a_{n-3}a_{n}, -a_{n-2}a_{n+1}-a_{1}a_{n}, -a_{n-1}a_{n+2}-a_{2}a_{n+1}, -a_{3}a_{n+2}),\NR
	d_2 		&=& (a_{n-2}a_{n}+a_{1}a_{n+1}, a_{n-1}a_{n+1}+a_{2}a_{n+2}).
\end{array}\]

We believe that all of these systems are integrable. They are Hamiltonian systems with a Poisson matrix determined by the equations \Ref{Poisson_Matrix}. For $n$ even, the Poisson structure has two Casimirs and the traces of $L^i$ together with an extra constant of motion obtained by Moser's technique give the integrability of the system. For $n$ odd the system it has one Casimirs and the traces of the $L^i$ give enough first integrals to ensure the integrability of the system.
We illustrate this with two examples, one for $n=7$ and one for $n=8$.

\begin{example}
For $n=7$ the matrix $L$ is given by
\[\displaystyle 
\begin{pmatrix}
	0&a_{1}&0&0&a_{7}&0&0\NR
	a_{1}&0&a_{2}&0&0&a_{8}&0\NR
	0&a_{2}&0&a_{3}&0&0&a_{9}\NR
	0&0&a_{3}&0&a_{4}&0&0\NR
	a_{7}&0&0&a_{4}&0&a_{5}&0\NR
	0&a_{8}&0&0&a_{5}&0&a_{6}\NR
	0&0&a_{9}&0&0&a_{6}&0
\end{pmatrix},
\]
while $B$ is given by
\[
\displaystyle
\begin{pmatrix}
	0&0&a_{1}a_{2}&-a_{4}a_{7}&0&a_{1}a_{8}+a_{5}a_{7}&0\NR
	0&0&0&a_{2}a_{3}&-a_{1}a_{7}-a_{5}a_{8}&0&a_{2}a_{9}+a_{6}a_{8}\NR
	-a_{1}a_{2}&0&0&0&a_{3}a_{4}&-a_{2}a_{8}-a_{6}a_{9}&0\NR
	a_{4}a_{7}&-a_{2}a_{3}&0&0&0&a_{4}a_{5}&-a_{3}a_{9}\NR
	0&a_{1}a_{7}+a_{5}a_{8}&-a_{3}a_{4}&0&0&0&a_{5}a_{6}\NR
	-a_{1}a_{8}-a_{5}a_{7}&0&a_{2}a_{8}+a_{6}a_{9}&-a_{4}a_{5}&0&0&0\NR
	0&-a_{2}a_{9}-a_{6}a_{8}&0&a_{3}a_{9}&-a_{5}a_{6}&0&0
\end{pmatrix}.
\]
The poisson bracket $\{\,,\}$ is determined by the following Poisson matrix.
\[\displaystyle \Gp= 
\begin{pmatrix}
	0&a_{1}a_{2}&0&0&0&0&-a_{1}a_{7}&a_{1}a_{8}&0\NR
	-a_{1}a_{2}&0&a_{2}a_{3}&0&0&0&0&-a_{2}a_{8}&a_{2}a_{9}\NR
	0&-a_{2}a_{3}&0&a_{3}a_{4}&0&0&0&0&-a_{3}a_{9}\NR
	0&0&-a_{3}a_{4}&0&a_{4}a_{5}&0&a_{4}a_{7}&0&0\NR
	0&0&0&-a_{4}a_{5}&0&a_{5}a_{6}&-a_{5}a_{7}&a_{5}a_{8}&0\NR
	0&0&0&0&-a_{5}a_{6}&0&0&-a_{6}a_{8}&a_{6}a_{9}\NR
	a_{1}a_{7}&0&0&-a_{4}a_{7}&a_{5}a_{7}&0&0&2\,a_{1}a_{5}&0\NR
	-a_{1}a_{8}&a_{2}a_{8}&0&0&-a_{5}a_{8}&a_{6}a_{8}&-2\,a_{1}a_{5}&0&2\,a_{2}a_{6}\NR
	0&-a_{2}a_{9}&a_{3}a_{9}&0&0&-a_{6}a_{9}&0&-2\,a_{2}a_{6}&0
\end{pmatrix}.
\]
The above system is equivalent to the following equations of motion.
\begin{eqnarray*}
	\dot{a}_{1} &=&a_{{1}}a_{{2}}^{2}+a_{{1}}a_{{8}}^{2}-a_{{1}}a_{{7}}^{2},\NR
	\dot{a}_{2} &=&a_{{2}}a_{{3}}^{2}-a_{{1}}^{2}a_{{2}}+a_{{2}}a_{{9}}^{2}-a_{{2}}a_{{8}}^{2},\NR
	\dot{a}_{3} &=&a_{{3}}a_{{4}}^{2}-a_{{2}}^{2}a_{{3}}-a_{{3}}a_{{9}}^{2},\NR
	\dot{a}_{4} &=&a_{{4}}a_{{5}}^{2}-a_{{3}}^{2}a_{{4}}+a_{{4}}a_{{7}}^{2},\NR
	\dot{a}_{5} &=&a_{{5}}a_{{6}}^{2}-a_{{4}}^{2}a_{{5}}+a_{{5}}a_{{8}}^{2}-a_{{5}}a_{{7}}^{2},\NR
	\dot{a}_{6} &=&-a_{{5}}^{2}a_{{6}}+a_{{6}}a_{{9}}^{2}-a_{{6}}a_{{8}}^{2},\NR
	\dot{a}_{7} &=&a_{{1}}^{2}a_{{7}}+a_{{5}}^{2}a_{{7}}-a_{{4}}^{2}a_{{7}}+2\,a_{{1}}a_{{5}}a_{{8}},\NR
	\dot{a}_{8} &=&a_2^2a_8-a_1^2a_8+a_6^2a_8-a_5^2a_8+2\,a_2a_6a_9-2\,a_1a_5a_7,\NR
	\dot{a}_{9} &=&a_3^2a_9-a_2^2a_9-a_6^2a_9-2\,a_2a_6a_8.
\end{eqnarray*}
The Casimir for the Poisson bracket is given by
	$$\det L = -2a_1a_3a_4a_6 (a_1a_5a_9 + a_2a_6a_7 - a_7a_8a_9).$$
Note that the constants of motion, $H_i=\tr L^i$ for $i=4,5,6$, together with the Hamiltonian $H_2 = \frac{1}{2}\left(a_1^2 + a_2^2 + \cdots + a_9^2\right)$ are functionally independent and in involution. Therefore the system is integrable.
\end{example}

\begin{example}
For n=8 the matrix $L$ is given by
\[\displaystyle L= \left(
\begin{array}{cccccccc}
	0&a_{1}&0&0&0&a_{8}&0&0\NR
 	a_{1}&0&a_{2}&0&0&0&a_{9}&0\NR
	0&a_{2}&0&a_{3}&0&0&0&a_{10}\NR
	0&0&a_{3}&0&a_{4}&0&0&0\NR
	0&0&0&a_{4}&0&a_{5}&0&0\NR
	a_{8}&0&0&0&a_{5}&0&a_{6}&0\NR
	0&a_{9}&0&0&0&a_{6}&0&a_{7}\NR
	0&0&a_{10}&0&0&0&a_{7}&0
\end{array}
\right),
\]
and the matrix $B$ is determined by the relations \Ref{B_Matrix_diagonals}.
The corresponding system is given by:
\begin{eqnarray*}
	\dot{a}_1 &=&a_{1}a_{2}^{2}+a_{1}a_{9}^{2}-a_{1}a_{8}^{2},\NR
	\dot{a}_2 &=&a_{2}a_{3}^{2}-a_{1}^{2}a_{2}+a_{2}a_{10}^{2}-a_{2}a_{9}^{2},\NR
	\dot{a}_3 &=&a_{3}a_{4}^{2}-a_{2}^{2}a_{3}-a_{3}a_{10}^{2},\NR
	\dot{a}_4 &=&a_{4}a_{5}^{2}-a_{3}^{2}a_{4},\NR
	\dot{a}_5 &=&a_{5}a_{6}^{2}-a_{4}^{2}a_{5}+a_{5}a_{8}^{2},\NR
	\dot{a}_6 &=&a_{6}a_{7}^{2}-a_{5}^{2}a_{6}+a_{6}a_{9}^{2}-a_{6}a_{8}^{2},\NR
	\dot{a}_7 &=&-a_{6}^{2}a_{7}+a_{7}a_{10}^{2}-a_{7}a_{9}^{2},\NR
	\dot{a}_8 &=&a_{1}^{2}a_{8}+a_{6}^{2}a_{8}-a_{5}^{2}a_{8}+2\,a_{1}a_{6}a_{9},\NR
	\dot{a}_9 &=&a_2^2a_9-a_1^2a_9+a_7^2a_9-a_6^2a_9+2\,a_2a_7a_{10}-2\,a_1a_6a_8,\NR
	\dot{a}_{10} &=&a_3^2a_{10}-a_2^2a_{10}-a_{7}^{2}a_{10}-2\,a_2a_7a_9.
\end{eqnarray*}
It is a Hamiltonian system with Poisson structure determined by the Poisson matrix
\[
\begin{pmatrix}
	0 &a_{1}a_{2} & 0 & 0 & 0 & 0 & 0 & -a_{1}a_{8} & a_{1}a_{9} & 0 \NR
	-a_{1}a_{2} &0&a_{2}a_{3} & 0 & 0 & 0 & 0 & 0 & -a_{2}a_{9} & a_{2}a_{10} \NR
	0 &-a_{2}a_{3} & 0 & a_{3}a_{4} & 0 & 0 & 0 & 0 & 0 & -a_{3}a_{10} \NR
	0 &0&-a_{3}a_{4} & 0 & a_{4}a_{5} & 0 & 0 & 0 & 0 & 0 \NR
	0 & 0 & 0 & -a_{4}a_{5} & 0 & a_{5}a_{6} & 0 & a_{5}a_{8} & 0 & 0 \NR
	0 & 0 & 0 & 0 & -a_{5}a_{6} & 0 & a_{6}a_{7} & -a_{6}a_{8} & a_{6}a_{9} & 0 \NR
	0 & 0 & 0 & 0 & 0 & -a_{6}a_{7} & 0 & 0 & -a_{7}a_{9} & a_{7}a_{10} \NR
	a_{1}a_{8} & 0 & 0 & 0 & -a_{5}a_{8} & a_{6}a_{8} & 0 & 0 & 2a_{1}a_{6} & 0 \NR
   -a_1a_9 & a_{2}a_{9} & 0 &  0 & 0 & -a_{6}a_{9}  & a_{7}a_{9} & -2a_{1}a_{6} & 0 & 2a_{2}a_{7} \NR
	0 & -a_{2}a_{10} & a_{3}a_{10} & 0 &  0 & 0 & -a_{7}a_{10} & 0 & -2a_{2}a_{7} & 0
\end{pmatrix}
\]
which has rank $8$.
The Hamiltonian of the system is $H_2 = \frac{1}{2}\left(a_1^2 + a_2^2 + \cdots + a_{10}^2\right)$. A 
constant of motion is obtained using Moser's technique.

\noindent
If we delete the odd numbered rows and columns of $L^2$ we get the matrix
$$
\La_o(L^2)=\begin{pmatrix}
a_1^{2}+a_2^{2}+a_9^{2}&a_2a_3&a_1a_8+a_6a_9&a_7a_9+a_2a_{10}\\
a_2a_3&a_3^2+a_4^2&a_4a_5&a_3a_{10}\\
a_1a_8+a_6a_9&a_4a_5&a_5^{2}+a_6^{2}+a_8^{2}&a_6a_7\\
a_7a_9+a_2a_{10}&a_3a_{10}&a_6a_7&a_7^2+a_{10}^2
\end{pmatrix}
 = 
\begin{pmatrix}
B_{{1}}&A_{{1}}&A_{{4}}&A_{{6}}\\
A_{{1}}&B_{{2}}&A_{{2}}&A_{{5}}\\
A_{{4}}&A_{{2}}&B_{{3}}&A_{{3}}\\ 
A_{{6}}&A_{{5}}&A_{{3}}&B_{{4}}
\end{pmatrix}.
$$
We have 
\begin{eqnarray*}
\dot{A_1}&=&\dot{(a_2a_3)}\\
&=&\dot{a_2}a_3+a_2\dot{a_3}=(a_2a_3^2+a_2a_{10}^2-a_1^2a_2-a_2a_9^2)a_3+a_2(a_3a_4^2-a_2^2a_3-a_3a_{10}^2)\\
&=&a_2a_3(a_3^2+a_4^2-a_1^2-a_2^2-a_9^2)\\
&=&A_1(B_2-B_1)
\end{eqnarray*}
and similarly the new variables $B_i,A_i$ satisfy the system
\begin{equation}
\label{equations}
\begin{array}{rcl}
\dot{B_1} &=& 2(A_1^2+A_6^2-A_4^2)\,,\\
\dot{B_2} &=& 2(A_2^2-A_1^2-A_5^2)\,,\\
\dot{B_3} &=& 2(A_3^2+A_4^2-A_2^2)\,,\\
\dot{B_4} &=& 2(A_5^2-A_3^2-A_6^2)\,,\\
\dot{A_1} &=& A_1(B_2-B_1)\,,\\
\dot{A_2} &=& A_2(B_3-B_2)\,,\\
\dot{A_3} &=& A_3(B_4-B_3)\,,\\
\dot{A_4} &=& A_4(B_1-B_3)+2A_3A_6\,,\\
\dot{A_5} &=& A_5(B_2-B_4)-2A_1A_6\,,\\
\dot{A_6} &=& A_6(B_4-B_1)-2A_3A_4+2A_1A_5\,.
\end{array}
\end{equation}

\noindent
This system can be written in Lax pair form $\dot{\La_o(L^2)}=\left[C,\La_o(L^2)\right]$ with
$$
C=
\begin{pmatrix}
 0       &  A_{{1}} &  -A_{{4}} &  A_{{6}}\NR
-A_{{1}} &    0     &  A_{{2}}  & -A_{{5}}\NR
 A_{{4}} & -A_{{2}} &     0     &  A_{{3}}\NR
-A_{{6}} &  A_{{5}} &  -A_{{3}} &     0
\end{pmatrix}.
$$
It is Hamiltonian with Hamiltonian function 
$$
H=\tr\left(\dfrac{\La_o\left(L^2\right)^2}{2}\right)=\dfrac{1}{2}(B_1^2+B_2^2+B_3^2+B_4^2)+A_1^2+A_2^2+A_3^2+A_4^2+A_5^2+A_6^2
$$ and Poisson matrix
$$
\begin{pmatrix}
0&0&0&0&A_1&0&0&-A_4&0&A_6\\
0&0&0&0&-A_1&A_2&0&0&-A_5&0\\
0&0&0&0&0&-A_2&A_3&A_4&0&0\\
0&0&0&0&0&0&-A_3&0&A_5&-A_6\\
-A_1&A_1&0&0&0&0&0&0&0&0\\
0&-A_2&A_2&0&0&0&0&0&0&0\\
0&0&-A_3&A_3&0&0&0&0&0&0\\
A_4&0&-A_4&0&0&0&0&0&0&A_3\\
0&A_5&0&-A_5&0&0&0&0&0&-A_1\\
-A_6&0&0&A_6&0&0&0&-A_3&A_1&0
\end{pmatrix}.
$$
It has 2 Casimir functions $B_1+B_2+B_3+B_4$ and $A_{{1}}A_{{2}}A_{{4}}+A_{{2}}A_{{3}}A_{{5}}$. The function
\[
\begin{split}
F&=A_{{1}}A_{{2}}A_{{4}}+A_{{2}}A_{{3}}A_{{5}}=
a_2a_3a_4a_5 \left( a_1a_8+a_6a_9\right) +a_3a_4a_5a_6a_7a_{10}=\\
&a_1a_2a_3a_4a_5a_8+a_2a_3a_4a_5a_6a_9+a_3a_4a_5a_6a_7a_{10}
\end{split}
\]
is a constant of motion for the original system. 
The integrals $H_2,H_4,H_6,F$ together with the two Casimirs given by
\[\displaystyle
\begin{array}{ccl}
	C_1 &=& a_1a_3a_5a_7,\\
	C_2 &=& \sqrt{\det L}-C_1 =a_{1}a_{4}a_{6}a_{10} + a_{2}a_{4}a_{7}a_{8} - a_{4}a_{8}a_{9}a_{10}.
\end{array}\]
ensure the integrability of the system.
\end{example}
In general for $n$ even Moser's technique gives the following additional constant of motion.
\renewcommand{\arraystretch}{1.5}
\begin{table}[h]
$$\begin{array}{|c|c|l|}
\hline
 n	& F=a_2a_3\ldots a_{n-3}(a_1a_n+a_{n-2}a_{n+1})+a_3a_4\ldots a_{n-1}a_{n+2}\\
\hline
      6		& a_1a_2a_3a_6+a_2a_3a_4a_7+a_3a_4a_5a_8\\
      8		& a_1a_2a_3a_4a_5a_8+a_2a_3a_4a_5a_6a_9+a_3a_4a_5a_6a_7a_{10}\\
     10	& a_1a_2a_3a_4a_5a_6a_7a_{10}+a_2a_3a_4a_5a_6a_7a_8a_{11}+a_3a_4a_5a_6a_7a_8a_9a_{12}\\
\hline
\end{array}$$
\caption{Additional constant of motion obtained using Moser's technique}
\end{table}

The following two tables contain the Casimirs of the Poisson structure for $m=3$.
\begin{table}[h]
$$\begin{array}{|c|c|l|}
\hline
 n	& C = -\frac{1}{2}\det L\\
\hline
      7		& a_1a_3a_4a_6							(a_1a_5a_9   +a_2a_6a_7		 - a_7a_8a_9)\\
      9		& a_1a_3a_4a_5a_6a_8					(a_1a_7a_{11}+a_2a_8a_9		 -a_9a_{10}a_{11})\\
     11	& a_1a_3a_4a_5a_6a_7a_8a_{10}			(a_1a_9a_{13}+a_2a_{10}a_{11}-a_{11}a_{12}a_{13})\\
     13	& a_1a_3a_4a_5a_6a_7a_8a_9a_{10}a_{12}	(a_1a_{11}a_{15}+a_2a_{12}a_{13}-a_{13}a_{14}a_{15})\\
\hline
     n		& a_1 a_3 a_4\cdots a_{n-3}a_{n-1}  (a_1a_{n-2}a_{n+2} + a_2a_{n-1}a_{n} - a_{n}a_{n+1}a_{n+2})\\
\hline
\end{array}$$
\caption{Casimirs for $m=3$ and $n$ odd}
\end{table}

\begin{table}[h]
$$\begin{array}{|c|l|l|}
\hline
n	& C_1						& C_2 = \sqrt{|\det{L}|} - C_1\\
\hline
 6		& a_1a_3a_5					& 							-(a_{1}a_{4}a_{8} + a_{2}a_{5}a_{6} - a_{6}a_{7}a_{8})\\
 8		& a_1a_3a_5a_7				& a_{4}						(a_{1}a_{6}a_{10} + a_{2}a_{7}a_{8} - a_{8}a_{9}a_{10})\\
 10		&  a_1a_3a_5a_7a_9			& -a_{4}a_{6}				(a_{1}a_{8}a_{12} + a_{2}a_{9}a_{10} - a_{10}a_{11}a_{12})\\
 12	    & a_1a_3a_5a_7a_9a_{11}		& a_{4}a_{6}a_{8}			(a_{1}a_{10}a_{14} + a_{2}a_{11}a_{12} - a_{12}a_{13}a_{14})\\
 14	    & a_1a_3a_5a_7a_9a_{11}a_{13}& -a_{4}a_{6}a_{8}a_{10}	(a_{1}a_{12}a_{16} + a_{2}a_{13}a_{14} - a_{14}a_{15}a_{16})\\
\hline
 n		& a_1 a_3 \cdots a_{n-3} a_{n-1}& a_4 a_6 \cdots a_{n-6} a_{n-4} (a_1a_{n-2}a_{n+2} + a_2a_{n-1}a_{n} - a_{n}a_{n+1}a_{n+2})\\
\hline
\end{array} $$
\caption{Casimirs for $m=3$ and $n$ even}
\end{table}

{\bf Acknowledgments}. The first  author  was supported by  a University of Cyprus Postdoctoral fellowship.
The work of the third  author  was co-funded by the European Regional Development Fund and the Republic of
Cyprus through the Research Promotion Foundation (Project: PENEK/0311/30).

\end{document}